\documentstyle[12pt,epsf]{article}
\def\appendix#1{
  \addtocounter{section}{1}
  \setcounter{equation}{0}
  \renewcommand{\thesection}{\Alph{section}}
 \section*{Appendix \thesection\protect\indent \parbox[t]{11.715cm} {#1}}
  \addcontentsline{toc}{section}{Appendix \thesection\ \ \ #1}
  }

\newcommand{\newsection}{
\setcounter{equation}{0}
\section}

\def\bea{\begin{eqnarray}}
\def\eea{\end{eqnarray}}
\def\be{\begin{equation}}
\def\ee{\end{equation}}

\newcommand{\Tr}[1]{\:{\rm Tr}\,#1}
\def\const{{\rm const}}
\def\e{{\,\rm e}\,}
\def\vac{|0\rangle}
\def\ps{|\Psi\rangle}
\def\pl{\langle\Psi |}
\def\Fd{\Phi^\dagger}
\def\Pd{\Pi^\dagger}
\def\d{\partial}
\def\D{\delta}
\def\f{\varphi}
\def\ka{\kappa}
\def\dd{^{\dagger}}
\renewcommand{\em}[1]{\varepsilon_{-\, #1 }}
\newcommand{\ep}[1]{\varepsilon_{+\, #1 }}
\def\es{\varepsilon}
\newcommand{\epm}[1]{\varepsilon_{\pm\, #1 }}
\newcommand{\psm}[1]{\psi_{-\, #1 }}
\newcommand{\psp}[1]{\psi_{+\, #1 }}
\newcommand{\pspm}[1]{\psi_{\pm\, #1 }}

\newcommand{\bpsm}[1]{\bar{\psi}_{-\, #1 }}
\newcommand{\bpsp}[1]{\bar{\psi}_{+\, #1 }}
\newcommand{\bpspm}[1]{\bar{\psi}_{\pm\, #1 }}
\newcommand{\am}[1]{a_{-\, #1 }}
\newcommand{\ap}[1]{a_{+\, #1 }}
\newcommand{\apm}[1]{a_{\pm\, #1 }}
\newcommand{\mn}[1]{\left\langle #1 \right\rangle}
\newcommand{\br}[1]{\left( #1 \right)}
\newcommand{\nor}[1]{\langle #1 | #1 \rangle}
\newcommand{\rf}[1]{(\ref{#1})}
\newcommand{\non}{\nonumber \\*}
\begin{document}
\thispagestyle{empty}
\begin{flushright}
NORDITA--HEP--98/63\\
ITEP--TH--72/98\\
\end{flushright}

\vskip 2true  cm

\begin{center}
{\Large\bf Screening of Fractional Charges}\\
\vskip 0.5cm
{\Large\bf in (2+1)-dimensional QED}
\vskip 1.5true cm

{\large\bf
Dmitri Diakonov$^{\diamond *}$ and Konstantin Zarembo$^{\dagger +}$}
\\
\vskip 1true cm
\noindent
{\it
$^\diamond $NORDITA, Blegdamsvej 17, 2100 Copenhagen \O, Denmark \\
\vskip .2true cm
$^* $Petersburg Nuclear Physics Institute, Gatchina,
St.Petersburg 188 350, Russia\\
\vskip .2true cm
$^\dagger $Department of Physics and Astronomy,
University of British Columbia,
 6224 Agricultural Road, Vancouver, B.C. Canada V6T 1Z1
\vskip .2true cm
$^+ $Institute of Theoretical and Experimental Physics,
 B. Cheremushkinskaya 25, 117259 Moscow, Russia}
\vskip 1.5cm
E-mails: {\tt diakonov@nordita.dk, zarembo@theory.physics.ubc.ca/@itep.ru}

\end{center}

\vskip 2true cm
\begin{abstract}
\noindent

We show that the logarithmically rising static potential between
opposite-charged sources in two dimensions is screened by dynamical
fields even if the probe charges are fractional, in units of the
charge of the dynamical fields. The effect is due to quantum mechanics:
the wave functions of the screening charges 
are superpositions of two bumps localized
both near the opposite- and the same-charge sources, so that each of
them gets exactly screened.

\end{abstract}
\newpage

\newsection{Introduction}

The static potential between trial external charges, or the Wilson
loop expectation value, carries important information about infrared
behavior of gauge theories. An infinite growth of the potential
provides the simplest criterium for confinement. However, in certain
theories the rising potential can be screened at large distances by
dynamical fields. The screening is inevitable if dynamical charges can
form neutral bound states with external sources, for example, when the
external and the dynamical charges belong to the same
representation of the gauge group (have the same magnitude in the
Abelian case). Since the potential between bare charges can be
arbitrary large, at some point the creation of a pair from the
vacuum becomes energetically favorable.  Each of the created charges
couples to the static charge of the opposite sign.  The interaction
energy between resulting bound states no more grows with the
separation.

The question of whether {\it fractional} charges can be screened or not
is more involved. This problem has been studied in two dimensions,
both in Abelian \cite{CJS71,GKMS95,ada96} and in non-Abelian
\cite{GKMS95,YM2} models. It appears that massless matter fields can
screen any Abelian fractional charge \cite{CJS71,GKMS95}.  In
non-Abelian case, massless fields in any representation of the gauge
group screen sources in the fundamental representation
\cite{GKMS95,YM2}, as follows from comparison of the 
models with massless
adjoint matter and with multiple flavors of fundamental matter
\cite{KS95}.

In this paper we consider the problem of charge screening in
three-dimensional scalar QED. This theory is confining when matter
decouples, as the Coulomb potential in two dimensions
grows logarithmically with distance. We
consider bosonic theory to purify the discussion,
because fermions, at least massive,
screen any charge in 2+1 dimensions \cite{AB97}. Three-dimensional
fermions induce the topological mass for a photon
at one loop \cite{CS} thus changing the logarithmically rising Coulomb
potential to the exponentially decreasing Yukawa one.
This phenomenon does not take place in scalar
QED, in which the photon remains massless.

We argue that, nevertheless,
any fractional static charge in 3D scalar QED is
screened, at least in the weak coupling regime. The
coupling constant $e^2$ in three dimensions has the dimension of mass,
so the weak coupling means that the ratio $e^2/m$ is small.
We do not expect any abrupt changes to happen as this parameter
is increased. Therefore, the screening, most probably, persists in the
strongly coupled theory as well.

We consider the static potential between external charges of the
magnitude $q$ separated by a distance $L$ (we imply that $0<q<1$
for simplicity
-- the screening of the integral part of the charge is obvious).
The  screening mechanism is very simple and
is based on the consideration
of a two-particle state. If
$q^2e^2\ln(L\mu)>2m$, the interaction energy is sufficiently
large to create a pair
of dynamical charges from the vacuum.
The parameter $\mu$ is determined by the typical scale of the screening
charge distribution, actually $\mu\sim\sqrt{me^2}$.
At first sight, such rearrangement
of the vacuum can only change $q$ to $q-1$,
but cannot stop the
logarithmic growth of the potential with $L$, since the dynamical and
the external charges form a bound state which is also charged. However,
the wave functions of dynamical particles need not be localized
near the opposite-charged sources only.
Suppose the wave function
has the form of a superposition of two states well localized near
each of the static sources. Let these
bumps be normalized to the probabilities $p$ and $1-p$,
respectively.
\begin{figure}[t]
\hspace*{5cm}
\epsfxsize=7cm
\epsfbox{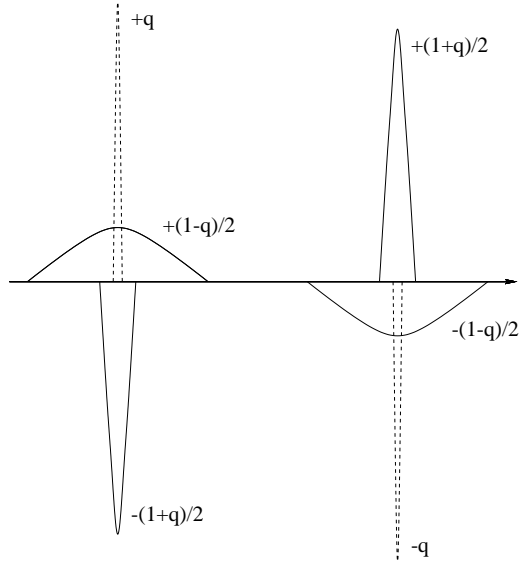}
\caption[x]{The delta-peaked external charges
(dashed lines)
together with the distribution of screening charges
(solid lines).}
\label{wvfig}
\end{figure}

This situation is schematically illustrated in fig.~\ref{wvfig}.
It is clear that if $p-(1-p)=q$, that is, if $p=(1+q)/2$,
the net charge localized near each of the external sources sums up
to zero. At the same time the total probabilities to find the
positive- and the negative-charged particles are both unities.
Only short-range interactions are present in such configuration
of charges; its energy does not grow with $L$, and,
thus, it becomes energetically favorable at very large separations. We
will argue that leaking of the charge to the region where it is
classically repulsed actually takes place for the Klein-Gordon particle
in the electric field of well-separated static charges. After that we
show that the configuration described above has smaller energy than
bare sources for sufficiently large $L$. The stability of this
configuration can be heuristically explained as follows. The
logarithmic Coulomb potential of the point charge is singular at short
distances, so it is quite natural that the negatively charged particle
form a bound state with the positive external source. This bound state
is described  by a larger part of the double-bump wave function of type
plotted in fig.~\ref{wvfig}. The charge of this bound state is
$-(1-q)/2$, so, as a whole, it attracts the positively charged
particle, which explains the
stability of the smaller part of the wave function.

The paper is organized as follows. In Sec.~2 we diagonalize the
Hamiltonian of the scalar QED in the presence of external charged
sources in the weak coupling approximation. In Sec.~3 we discuss
the Klein-Gordon equation for a charged particle in the electric
field of a dipole.
In Sec.~4 the two-particle state becoming energetically
favorable at large separation is considered in more detail.
In Sec.~5 and in Appendix we comment on the path integral for
scalar QED in the presence of external charges.

\newsection{External charges in scalar QED}\label{hsec}

We consider scalar QED in three dimensions. The Lagrangian density
of this theory is
\be
{\cal L}=-\frac14\,F_{\mu\nu}F^{\mu\nu}+(D_\mu\Phi)^\dagger D^\mu\Phi
-m^2\Fd\Phi,
\ee
where $\Phi$ is a complex scalar field and
\be
D_\mu=\d_\mu+ieA_\mu.
\ee
For our purposes the canonical formalism is more appropriate.

In the Schr\"odinger representation,  $A_0=0$
and  $E_i=F_{0i}=\dot{A}_i$, $A_i$, $\Pi=\dot{\Phi}^\dagger$,
$\Phi$, $\Pd=\dot{\Phi}$,
$\Fd$ are canonical variables:
\be
[A_i(x),E_j(y)]=i\D_{ij}\D(x-y),
\ee
\be
[\Phi(x),\Pi(y)]=i\D(x-y)=[\Fd(x),\Pd(y)].
\ee
The Hamiltonian is
\be\label{ham}
H=\int d^2x\,\left[\frac12\,E_i^2+\frac14\,F_{ij}^2+\Pd\Pi
+(D_i\Phi)^\dagger D_i\Phi+m^2\Fd\Phi\right].
\ee
The physical states
in the presence of external charges are
 subject to the Gauss' law constraint:
\be\label{gl}
\d_iE_i|\Psi_{\rm phys}\rangle=(J_0+\rho)|\Psi_{\rm phys}\rangle,
\ee
where $J_0$ is the charge density operator,
\be
J_0(x)=ie\Bigl(\Fd(x)\Pd(x)-\Phi(x)\Pi(x)\Bigr)
\ee
and $\rho$ is the density of external sources. In our case of two
well-separated point-like charges
\be\label{defrho}
\rho(x)=qe\Bigl(\D\br{x+L/2}-\D\br{x- L/2}\Bigr).
\ee
The potential of interaction between charges is equal to the difference
of the ground state energies in the sectors of the Hilbert space
defined by the Gauss' law with and without external sources.

The qualitative picture of charge screening is based on purely
classical notion of electric field
which becomes strong enough to create
pairs. It is difficult to visualize this picture in the Hamiltonian
formalism, where electric fields are operators in the Hilbert space
and logarithmically rising electrostatic potentials are somehow
encoded in the dependence of the wave functional on $A_i$.
However, it is possible to introduce classical, c-number
fields which play the role of electric potentials in the Hamiltonian
formalism, despite $A_0=0$
by definition.

Consider the following Hamiltonian:
\be\label{hamf}
H(\f)=H+\int d^2x\,\f\br{\d_iE_i-J_0-\rho},
\ee
which now acts in the unconstraint Hilbert space. The
eigenfunctions $\ps$ and the eigenvalues $E$ of this Hamiltonian
are the functionals of $\f$:
\be\label{sef}
H(\f)\ps=E\ps.
\ee
We want to show that, if $\f$ is determined by
stationarity condition
\be\label{sta}
\frac{\D E}{\D \f}=0,
\ee
the state $\ps$ satisfies the Gauss' law \rf{gl} and is an eigenstate of
the Hamiltonian \rf{ham} with the eigenvalue $E$.

Both of the Hamiltonians \rf{ham} and
\rf{hamf} commute with the Gauss' law and with one another,
so they can be
simultaneously diagonalized. Therefore, it is sufficient to show that
the Gauss' law is satisfied in average. This immediately follows from
the stationarity condition \rf{sta} and
the Schr\"odinger equation \rf{sef}:
$$
0=
\frac{\D}{\D\f}\,\frac{\pl H(\f)\ps}{\nor{\Psi}}
=2\,\frac{\pl (H(\f)-E)\,\frac{\D}{\D \f}\ps}{\nor{\Psi}}
+\frac{\pl\frac{\D H(\f)}{\D\f}\ps}{\nor{\Psi}}
=\frac{\pl \d_iE_i-J_0-\rho\ps}{\nor{\Psi}}.
$$

The interaction of the scalar fields with photons entering the Hamiltonian
through the covariant derivative squared term can be disregarded in the
weak coupling limit since it is
of order $e^2/m$ and is not enhanced by a $\ln(L\mu)$
factor. The Hamiltonian $H(\f)$ is quadratic in this approximation:
\bea\label{hamq}
H(\f)&=&\int d^2x\,\left[\frac12\,E_i^2-E_i\d_i\f+\frac14\,F_{ij}^2
\right.\non
&&\left.+\Pd\Pi
+ie\f(\Phi\Pi-\Fd\Pd)+\d_i\Fd \d_i\Phi+m^2\Fd\Phi-\f\rho\right],
\eea
and can be explicitly diagonalized.

Solutions of the Schr\"odinger equation for the Hamiltonian
\rf{hamq} have factorized form
\be
\Psi=\Psi_{\rm gauge}[A]\Psi_{\rm matter}[\Phi,\Fd].
\ee
We first consider the gauge-field part of the wave function. The ground
state is described by a Gaussian wave functional:
\be
\Psi_{\rm gauge}[A]=\exp\left(-\frac12\int d^2xd^2y\,A_i(x)K_{ij}(x,y)A_j(y)
+i\int d^2x\,{\cal E}_i(x)A_i(x) \right).
\ee
Substituting this expression in the Schr\"odinger equation (electric
fields act on the wave functional as variational derivatives:
$E_i=-i\D/\D A_i$), we obtain for $K_{ij}$ and ${\cal E}_i$:
\bea
&&K_{ij}=\sqrt{-\d^2}\br{\D_{ij}+\frac{\d_i\d_j}{-\d^2}},
\\*
&&{\cal E}_i=\d_i\f.
\eea
The energy of this state is
\be\label{ea}
E_{\rm gauge}=E_0-\frac12\int d^2x\,(\d \f)^2,
\ee
where $E_0$ is the divergent zero-point energy
$$
E_0=\frac12\Tr K=\int \frac{d^2p}{(2\pi)^2}\,p,
$$
which does not depend on $\f$ and is omitted below.

The Hamiltonian for matter fields can be diagonalized introducing
creation and annihilation operators:
\be\label{cran}
[H(\f),a\dd]=\varepsilon a\dd.
\ee
Since Hamiltonian commutes with electric charge
\be
Q=\int d^2x\,J_0,
\ee
operators satisfying eq.~\rf{cran} are linear combinations of
$\Pi$ and $\Fd$ (or of $\Pd$ and $\Phi$):
\be
a\dd=\int d^2x\,\br{\Pi\psi+\Fd\tilde{\psi}}.
\ee
Substituting this operator in eq.~\rf{cran} we find that
$\tilde{\psi}=i(\varepsilon+e\f)\psi$, where $\psi$ and $\varepsilon$
are determined by the equation
\be\label{kg}
\left[\br{\epm{n}+e\f}^2+\d^2-m^2\right]\pspm{n}=0.
\ee
Here the subscripts $\pm$ mark positive- and negative-energy states:
\be
\ep{n}>0,~~~~\em{n}<0.
\ee
The equality \rf{kg} is nothing but the
 Klein-Gordon equation for  eigenmodes
in the time-in\-de\-pen\-dent
external field $A_\mu=\D_{\mu 0}\f$. Its solutions
form two complete sets of functions normalized by \cite{mig72}
\bea\label{norma}
\int d^2x\,\bpspm{m}\br{\epm{m}+\epm{n}+2e\f}\pspm{n}&=&\pm\D_{mn},
\\*
\int d^2x\,\bpspm{m}\br{\epm{m}+
\varepsilon_{\mp\,n}+2e\f}\psi_{\mp\,n}&=&0.
\eea

These eigenfunctions determine two sets of operators
\bea
\ap{n}\dd&=&\int d^2x\,\left[\Pi(x)\psp{n}(x)
                +i\Fd(x)\br{\ep{n}+e\f(x)}\psp{n}(x)\right],
\non
\ap{n}&=&\int d^2x\,\left[\bpsp{n}(x)\Pd(x)
             -i\bpsp n(x)\br{\ep{n}+e\f(x)}\Phi(x)\right],
\non
\am{n}\dd&=&\int d^2x\,\left[\bpsm{n}(x)\Pd(x)
             -i\bpsm n(x)\br{\em{n}+e\f(x)}\Phi(x)\right],
\non
\am{n}&=&\int d^2x\,\left[\Pi(x)\psm{n}(x)
                +i\Fd(x)\br{\em{n}+e\f(x)}\psm{n}(x)\right],
\eea
which create and annihilate
particles of charge $\pm e$ and energy $|\epm{n}|$:
\be
[Q,\apm{n}\dd]=\pm e\apm{n}\dd,~~~~[Q,\apm{n}]=\mp e\apm{n}
\ee
\be
[H(\f),\apm{n}\dd]=\pm\epm{n}\apm{n}\dd=|\epm{n}|\apm{n}\dd,
~~~~[H(\f),\apm{n}]=\mp\epm{n}\apm{n}=-|\epm{n}|\apm{n},
\ee
\be
[\apm n,\apm m\dd]=\D_{nm},~~~~[\apm n,\apm m]=0=[\apm n,a_{\mp\, m}\dd].
\ee
The field variables are expressed in terms of creation and annihilation
operators as
\bea
\Phi(x)&=&i\sum_n\br{\psp{n}(x)\ap{n}-\psm{n}(x)\am{n}\dd},
\non
\Pi(x)&=&\sum_n\left[\ap{n}\dd\bpsp{n}(x)\br{\ep{n}+e\f(x)}
-\am{n}\bpsm{n}(x)\br{\em{n}+e\f(x)}\right].
\eea

The total energy is comprised of $E_{\rm gauge}$ given by eq.~\rf{ea}, the
energy of the matter fields, $E_{\rm matter}$, and the source term:
\be\label{ef}
E=-\int d^2x\,\left[\frac12\,(\d\f)^2+\f\rho\right]+E_{\rm matter}.
\ee
The energy of the state satisfying the Gauss' law corresponds to the
extremum of this functional, which is determined
by the following equation:
\be\label{pois}
-\d^2\f+\rho+\mn{J_0}=0,
\ee
where we used the fact that $\D E_{\rm matter}/\D\f
=\mn{\D H_{\rm matter}(\f)/\D\f}=
-\mn{J_0}$. Note that this extremum is a maximum of $E(\f)$.

When the separation of external charges is not very large,
an empty state,
\be
\apm n\vac=0,
\ee
has the lowest energy. The vacuum charge density can be expanded in
powers of $1/m$:
\be\label{avj0}
\langle 0|J_0\vac=\const\,\frac{e^2}{m}\,\d^2\f+\ldots,
\ee
and is small compared to the first term in eq.~\rf{pois}.
Therefore, we can safely omit vacuum
contributions to the energy and to the
charge density.

The equation \rf{pois} is then solved by
\be\label{logfrac}
\f=\frac{eq}{2\pi}\,\ln \frac{|x+L/2|}{|x-L/2|}
\ee
and \rf{ef} gives
the Coulomb law:
\be\label{coul}
E=\frac{q^2e^2}{2\pi}\,\ln\frac{L}{2\zeta}.
\ee
Here $\zeta$ is an UV cutoff necessary to regularize an infinite
Coulomb self-energy of the static charges (see sec.~4 for the precise
definition).
However, the logarithmic raise of the interaction
energy cannot last infinitely without a substantial rearrangement of
the vacuum. To study it, we consider next the spectrum of the
Klein-Gordon equation for the potential \rf{logfrac} of the electric
dipole.

\newsection{Klein-Gordon equation in the dipole field}\label{kgsec}

First, when the distance
$L$ is small, the Klein-Gordon equation \rf{kg}
has no normalizable solutions and its
spectrum consists of two continua with $\ep{}>m$ and with
$\em{}<-m$. Near the boundaries of the spectrum, for $\epm{}=
\pm(m+\lambda)$, $\lambda\ll m$, the
non-relativistic approximation can be used: the
first term in eq.~\rf{kg} can be expanded in $e\f$, and
the eigenfunctions $\pspm{}$ satisfy ordinary Schr\"odinger
equation with the potential $\mp e\f$:

\be\label{nrel}
\br{-\frac{1}{2m}\,\d^2\mp e\f-\lambda_\pm}
\pspm{}=0.
\ee
The potential energy for positively charged (coming from the upper
\begin{figure}[t]
\hspace*{5cm}
\epsfxsize=7cm
\epsfbox{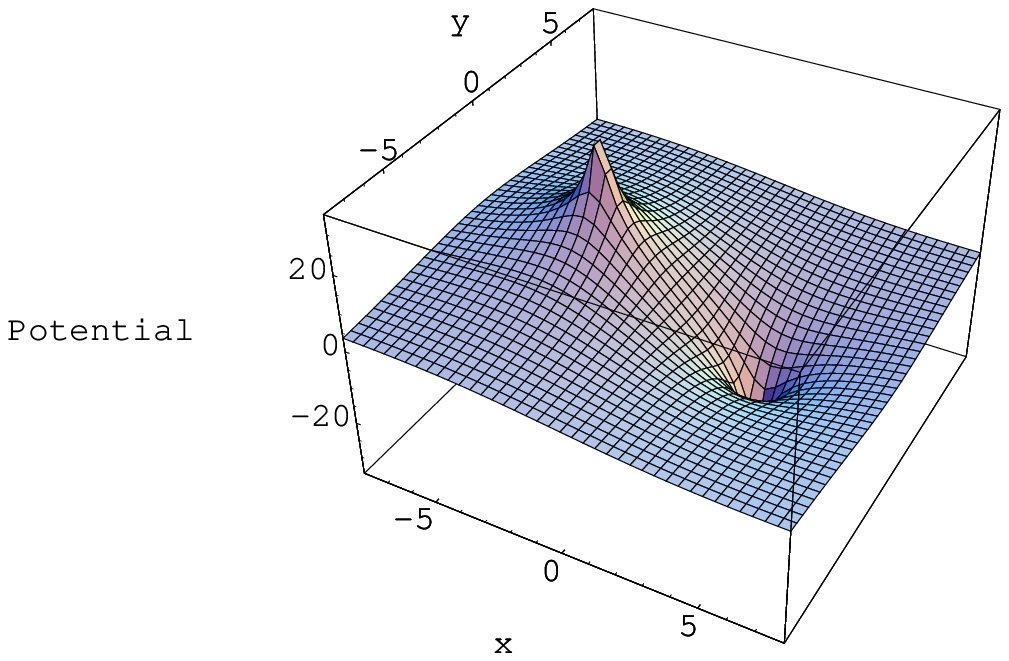}
\caption[x]{The potential \rf{logfrac}.}
\label{logfig}
\end{figure}
continuum) particles, shown in fig.~\ref{logfig}, has a form
of the separated peak and the well. For sufficiently
large $L$ the attraction by the well
becomes strong enough for a discrete level to appear.
Since a positive charge is
attracted to the source at $x=L/2$ and is repulsed from the one at
$x=-L/2$, its wave function $\psp{0}$ is localized near $x=L/2$. The
wave function of the negative charge is localized near $x=-L/2$ and its
energy is $\em 0=-\ep 0$ by symmetry. The lowest positive- and
negative-energy levels converge with the increase of $L$ and collide at
zero for some critical
\begin{figure}[t] \hspace*{5cm} \epsfxsize=7cm
\epsfbox{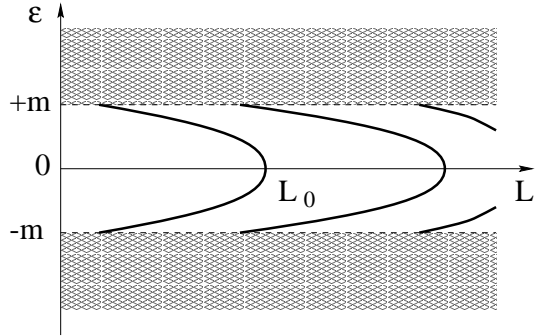}
\caption[x]{Spectrum of the Klein-Gordon equation for the potential
\protect\rf{logfrac}.}
\label{specfig}
\end{figure}
value of $L=L_0$ (fig.~\ref{specfig}). After that they do not
disappear, but rather go off to the complex plane \cite{zp71}. This
behavior of the eigenvalues is generic for Klein-Gordon equation in
strong electric fields \cite{mig72,zp71,GMR85}. For $L>L_0$ vacuum
polarization can no longer be neglected and the external electric field
creates a pair of charged particles in the vacuum.  This effect is
analogous to a pair creation in the field of a heavy ion
with the nuclear charge $Z>137$ \cite{zp71}.

For large $L$ the ground state wave function is no longer localized
near the well of the potential. Rather, the wave function has the shape
with two bumps, like the one used in the qualitative
arguments in the introduction.
As $L$ grows, the charge leaks from the
well to the region where the potential is peaked. This, at first sight,
anti-intuitive behavior reflects a generic property of
a charged Klein-Gordon particle to form a bound state in a sufficiently
strong repulsive electric potential \cite{zp71,mig72,GMR85}. The fact that
the charge is redistributed between the well and the
peak of the potential can be proved by the following arguments. For
$L=L_0$ both eigenvalues $\em 0$ and $\ep 0$ are equal to zero and the
Klein-Gordon equation formally has the form of the
Schr\"odinger one:
\be
\left[-\d^2-\br{\frac{e^2q}{2\pi}\,\ln \frac{|x+L/2|}{|x-L/2|}}^2
+m^2\right]\pspm 0=0~~~~(L=L_0).
\ee
The potential here has the shape of a symmetric double well. The
ground state wave function is symmetrically distributed between the
two wells. By continuity reasons, as $L$ is decreased, the charge
begin to leak from the region near the repulsive source to the
attractive one, and eventually all the positive charge is concentrated
near $+L/2$ and the negative one near $-L/2$.

\newsection{Screening of the logarithmic potential}

So far, the vacuum sector was considered.
However, when $e^2\ln(L\mu)\sim m$, the two-particle state
\be\label{2}
| 2\rangle=\ap 0\dd\am 0\dd\vac
\ee
can become energetically more favorable, if the dynamical charges screen
the sources and reduce the Coulomb energy by an amount
sufficient to create a pair. The energy of the state \rf{2} is
\be
E_{\rm matter}=\ep 0-\em 0
\ee
and the induced charge density is
\be\label{j02}
\langle 2|J_0| 2 \rangle
=2e\bpsp 0(\ep 0+e\f)\psp 0+2e\bpsm 0(\em 0+e\f)\psm 0.
\ee
Vacuum contributions are neglected here.

The charged particles which cause the screening of the static charges
are non-relativistic, since the screened electric fields are small
everywhere, unlike the unscreened ones. Therefore, it is possible to
use the non-relativistic approximation \rf{nrel} to the Klein-Gordon
equation.  In this approximation, the wave functions $\pspm {}$ -- we
omit the subscript $0$ for brevity -- are normalized to $1/(2m)$, as
follows from equation \rf{norma}. For the sake of clarity it is,
however, convenient to introduce new wave functions, $\psi_\pm^\prime=
\sqrt{2m}\psi_\pm$, normalized to unity,
as it is custom in the non-relativistic limit:

\be\label{nor}
\int d^2x\,|\psi_\pm^\prime|^2=1,
\ee
whereas the induced charged density is
\be\label{j}
\mn{J_0}=e\br{|\psi_+^\prime|^2-|\psi_-^\prime|^2}.
\ee
The total energy of the two-particle state is
\be\label{en}
E=2m+\frac{1}{2m}\!
\int\!\! d^2x\,\Bigl(|\d\psi_+^\prime|^2+|\d\psi_-^\prime|^2\Bigr)
-\!\frac{1}{4\pi}\!\int\!\!d^2x d^2y\left[\rho(x)+\mn{J_0(x)}\right]
\ln|x-y|\left[\rho(y)+\mn{J_0(y)}\right].
\ee
This expression is obtained after $2m+\lambda_++\lambda_-$ is
substituted for $E_{{\rm matter}}$ in eq.~\rf{ef}, and the solution
of the Poisson equation \rf{pois} is substituted for $\f$.

The energy \rf{en} can be regarded as a functional of
$\psi_\pm^\prime$. It is straightforward to check
that the minimum of this functional is determined exactly by the
Schr\"odinger equation \rf{nrel}. The coupled set of equations
\rf{pois}, \rf{nrel}, \rf{nor} and \rf{j} constitute a rather
complicated eigenvalue problem. The ground state corresponds to the
global minimum, and can be found numerically. Instead of solving
these equations directly we will suggest simple variational wave
functions $\psi_\pm^\prime$ corresponding to a two-particle state
whose energy does not grow with the separation of the external charges
$L$.

We take the variational wave functions in a form of a superposition of
states with charges $\pm e$ localized in the vicinity of the sources
with charges $\pm eq,\;\;q<1$, see fig.~\ref{wvfig}. For symmetry
reasons we take $\psi_-^\prime(x)=\psi_+^\prime(-x)$:

\[
\psi_+^\prime(x)=\phi_1\left(x-\frac{L}{2}\right)
+\phi_2\left(x+\frac{L}{2}\right), \]
\be\label{an}
\psi_-^\prime(x)=\phi_2\left(x-\frac{L}{2}\right)
+\phi_1\left(x+\frac{L}{2}\right).
\ee
The functions $\phi_{1,2}$ are supposed to be well-localized
and normalized to

\be\label{sub}
\int d^2 x\, \phi_{1,2}^2=\frac{1\mp q}{2}.
\ee
For large $L$ the overlap between $\phi_1$ and $\phi_2$ is exponentially
small and can be neglected, therefore the wave functions \rf{an} are
normalized to unity. This ansatz corresponds to the distribution of
charges described in the introduction. The $(1+q)/2$ portion of the
dynamical charge is localized near the source of the opposite sign and
the $(1-q)/2$ portion is localized near the one with the same sign.

Neglecting the exponentially small overlap of $\phi_{1,2}$ we get
for the total charge density:

\[
\rho(x)+\mn{J_0(x)}
=e^2\left\{\left[q\,\delta\left(x+\frac{L}{2}\right)
+\phi_1^2\left(x+\frac{L}{2}\right)
-\phi_2^2\left(x+\frac{L}{2}\right)\right]\right.
\]
\be\label{fullcharge}
\left.
-\left[q\,\delta\left(x-\frac{L}{2}\right)
+\phi_1^2\left(x-\frac{L}{2}\right)
-\phi_2^2\left(x-\frac{L}{2}\right)\right]\right\}.
\ee
We see that the total charge density noticeably differs from zero only
in the vicinity of the points $x=-L/2$ and $x=+L/2$. The screening
of the delta-peaked external sources is achieved if the integral
of the total charge density over the region much smaller than $L$
is zero. This is guaranteed by our choice of the normalization
condition \rf{sub}.

To make an estimate of the minimal energy of the two-particle
screening state and, hence, of the critical separation between the
external charges where the rising potential breaks up, we take
a simple Gaussian ansatz for the wave functions $\phi_{1,2}$:

\be\label{Gauss}
\phi_{1,2}(x)=\sqrt{\frac{1\mp q}{4\pi a_{1,2}^2}}\exp\left(
-\frac{x^2}{4a_{1,2}^2}\right),
\ee
where the widths of the wave functions $a_{1,2}$ are the variational
parameters. To make all integrals finite we shall temporarily introduce
a Gaussian smearing of the external sources replacing

\be\label{smear}
q\delta(x)\rightarrow \frac{q}{2\pi\zeta^2}\exp\left(
-\frac{x^2}{2\zeta^2}\right),\;\;\;\;\zeta\rightarrow 0.
\ee

Now one has to substitute the trial wave functions \rf{Gauss}
into the energy functional \rf{en} and to find the best
widths $a_{1,2}$ from its minimum. The integrals are readily performed
by using the Fourier transforms. Recalling that the Fourier
transform of $\ln(x^2)/4\pi$ is $-1/k^2$ we get for the
interaction or the potential energy term in the total energy
(the last term in eq.\rf{en}):

\be\label{poten}
E_{pot}(L)=-e^2\int\frac{d^2k}{(2\pi)^2}\frac{e^{ik\cdot L}-1}{k^2}
\left[q\,e^{-\frac{\zeta^2k^2}{2}}+\frac{1-q}{2}e^{-\frac{a_1^2k^2}{2}}
-\frac{1+q}{2}e^{-\frac{a_2^2k^2}{2}}\right]^2.
\ee
The term proportional to $\exp(ik\cdot L)$ accounts for the interaction
between the regions near $x+ L/2$ and near $x-L/2$, while the
subtracted term proportional to unity takes into account the
self-interaction of charge distributions inside these regions.

Eq.\rf{poten} should be compared to the interaction of two bare
external charges:

\be\label{bare}
E_{bare}(L)=-e^2\int\frac{d^2k}{(2\pi)^2}\frac{e^{ik\cdot L}-1}{k^2}
\left[q\,e^{-\frac{\zeta^2k^2}{2}}\right]^2
=\frac{e^2q^2}{4\pi}\left(\ln\frac{L^2}{4\zeta^2}+\gamma_E\right).
\ee
Terms exponentially small in $L/\zeta$ have been neglected here.
The integral \rf{poten} is immediately calculated using \rf{bare},
yielding

\[
E_{pot}=\frac{e^2}{4\pi}\left[q^2\ln\frac{L^2}{4\zeta^2}
+\left(\frac{1-q}{2}\right)^2\ln\frac{L^2}{4a_1^2}
+\left(\frac{1+q}{2}\right)^2\ln\frac{L^2}{4a_2^2}\right.\]
\be\label{poten1}
\left.+2q\frac{1-q}{2}\ln\frac{L^2}{2(\zeta^2+a_1^2)}
-2q\frac{1+q}{2}\ln\frac{L^2}{2(\zeta^2+a_2^2)}
-2\frac{1-q}{2}\frac{1+q}{2}\ln\frac{L^2}{2(a_1^2+a_2^2)}\right].
\ee
The coefficient in front of $\ln L^2$ is zero, so that the energy is
now independent of $L$, up to exponentially small corrections which
are neglected. It is exactly the screening effect we are after,
and it is due to the choice of the normalization of charge
distributions, eq.\rf{sub}. Neglecting also the spread of the external
charges $\zeta^2$ as compared to $a_{1,2}^2$ we get

\[
E_{pot}=\frac{e^2}{4\pi}\frac{1}{4}\left[2(1-q)(1+q)\ln(a_1^2+a_2^2)
-(1-q)(1+3q)\ln a_1^2- (1+q)(1-3q)\ln a_2^2\right.\]
\be\label{poten2}
\left. -2(1+3q^2)\ln 2-4q^2\ln\zeta^2\right].
\ee

The kinetic energy term (the second term in eq.\rf{en}) is

\be\label{kin}
E_{kin}=\frac{1}{4m}\left(\frac{1-q}{a_1^2}+\frac{1+q}{a_2^2}\right)
\ee
The sum, $E_{kin}+E_{pot}$, has a minimum at

\be\label{min}
a_{1,2}^2=\frac{4\pi}{me^2}
\frac{1\pm q+4q^2+(1\pm q)\sqrt{1+8q^2}}{4q^2(1\mp q)},
\ee
which should be substituted into \rf{kin} and \rf{poten2} to
get a variational estimate of the energy of the two-particle state
screening the external charges. Naturally, the energy-at-rest, $2m$,
should be added, too.

It follows from \rf{min} that the distribution of the dynamical charge
having the {\it same} sign as the external charge ($a_1$) is
{\it broader} than that having the opposite charge ($a_2$). For
example, if the external charge is one half of the dynamical charge
($q=1/2$) the same-charge cloud is about 2.5 times broader than
the opposite-charge cloud. See the table, where examples for other
values of $q$ are given.

Finally, the energy of the two-particle ground state can be written as

\be\label{ground}
E=2m+\frac{e^2q^2}{4\pi}\left(\ln\frac{4\pi C_q^2}{me^2q^24\zeta^2}
+\gamma_E\right),
\ee
where $C_q$ is a number of the order of unity coming from substituting
the best values of $a_{1,2}$ given by eq.\rf{min} into
$E_{kin}+E_{pot}$. Notice that the dependence on the spread of the
delta-peaked external charges, $\zeta$, is the same as in the case
of the bare charges, eq.\rf{bare}. This is because the extended
dynamical charge distribution cannot screen the logarithmic potential
at small separations.

\begin{center}{\bf Table}
\vskip .5true cm

\begin{tabular}{|c|c|c|c|}
\hline
$q$ &  $a_1\sqrt{\frac{me^2}{4\pi}}$ & $a_2\sqrt{\frac{me^2}{4\pi}}$ &
$C_q$ \\
\hline
\hline
0.1 & 7.96 & 6.53 & 0.536 \\
\hline
$\frac{1}{3}$ & 3.49 & 1.85 & 0.589 \\
\hline
$\frac{1}{2}$ & 3.19 & 1.26 & 0.649 \\
\hline
$\frac{2}{3}$ & 3.44 & 0.976 & 0.717 \\
\hline
0.9 & 5.65 & 0.776 & 0.824 \\
\hline
\end{tabular}\end{center}
\vskip .1true cm

The logarithmically rising potential between external charges at
large separations breaks up when the bare energy \rf{bare} exceeds the
energy of the screening state, eq.\rf{ground}. It happens at the
critical separation between the external charges

\be\label{crit}
L_c=C_q\sqrt{\frac{4\pi}{me^2q^2}}\exp\left(\frac{4\pi m}
{e^2q^2}\right).
\ee

Since we assume the non-relativistic limit, $m\gg e^2/4\pi$, this
distance is exponentially large. Notice that the widths of the
screening distributions $a_{1,2}$ as given by eq.\rf{min} are
much less than the critical distance $L_c$, which justifies
neglecting of the overlaps between the screening clouds belonging
to the two centers.

The numerical values of the coefficient $C_q$ are given in the table
for certain values of $q$, together with the values of the widths
$a_{1,2}$ measured in natural units of $\sqrt{me^2/4\pi}$. Since we
have used a variational estimate for the ground state energy, the
true minimum can be only lower, that is to say that the numerical
coefficient $C_q$ of the order of unity in eqs.\rf{ground},\rf{crit}
can be somewhat smaller than given in the table. However, the
dependence on the algebraic parameters in these equations follow
from the dimension analysis, and is of a general nature.

At $L\approx L_c$ the logarithmic growth of the potential stops;
more precisely it slowly grows
approaching its asymptotic value at infinity \rf{ground}, the deviation
corresponding to the residual
%van der Waals-like
forces between neutral charge clouds at $x=\pm L$.

\newsection{Path-integral approach}

The arguments we used above are in essence the variational ones.
For this reason, we preferred to use the Hamiltonian formalism.
Although the discussion of the screening from the path-integral
point of view is beyond the scope of the present paper, we would like
to outline how the main ingredients of our analysis can be derived
from the path integral. Here we consider only the vacuum sector.

The vacuum average of the
Wilson loop infinitely stretched in the time direction, which determines
the energy of two static charges, is given by the path integral

\bea\label{zed}
Z&=&\int DAD\Fd D\Phi\,\exp\left\{
i\int d^4x\,\left[-\frac14\,F_{\mu\nu}F^{\mu\nu}+
(D_\mu\Phi)^\dagger D^\mu\Phi-m^2\Fd\Phi\right]
\right.\non
&&\left.+iqe\int dx^0\,\Bigl(A_0(x^0,-L/2)-A_0(x^0,+L/2)\Bigr)
\right\}.
\eea
Integration over the scalar fields induces an effective action
for the gauge potentials:
\be\label{ind}
\Delta S=i\Tr\ln(-D^2-m^2+i0).
\ee
The next step, justified by the smallness of the coupling, is to
calculate the remaining integral over $A_\mu$ in the saddle-point
approximation. This amounts to solving classical equations of motion
taking into account the induced action \rf{ind} and the source
term in \rf{zed}. We are going to show that these saddle-point
equations are nothing but the ones derived in
Secs.~\ref{hsec},~\ref{kgsec} in the Hamiltonian formalism.

Owing to the symmetries of the problem the classical fields
are time-independent, and $A_i^{\rm cl}=0$.
For $A_0^{\rm cl}$ we get the Poisson equation,

\be\label{sp}
-\d_i\d_iA_0^{\rm cl}+\rho+\mn{J_0}=0,
\ee
where $\rho$ is the same as in \rf{defrho} and the induced charge
density is
\be\label{chad}
\mn{J_0}=i\,\frac{\D}{\D A_0^{\rm cl}}\,\Tr\ln(-D^2-m^2+i0)
=e\left.\Bigl(\overrightarrow{D}_0G(x,y)
+G(x,y)\overleftarrow{D}_0\Bigr)\right|_{x=y},
\ee
where
\be\label{green}
G(x,y)=\mn{x\left|\frac{1}{-D^2-m^2+i0}\right|y}.
\ee
We leave the calculation of the induced charge density to Appendix,
where we show that it reduces to solving the Klein-Gordon
equation \rf{kg} with $\f$ replaced by $A_0^{\rm cl}$
and that the saddle-point equation \rf{sp} coincides with \rf{pois}.

\newsection{Conclusions}

To summarize, the infinitely rising potential between fractionally
charged external (pro\-be) sources is screened in (2+1)-dimensional
scalar QED.
%We have demonstrated it using both the Hamiltonian
%and path-integral approaches.
Of course, if the mass of the
dynamical fields is large, the rising potential persists at
intermediate scales. The critical distance is exponentially large in
$m/e^2$, in contrast to 3D spinor QED where the screening length is of
the order of $1/e^2$ \cite{AB97}.

The screening is a typical quantum-mechanical effect: the wave
functions of the screening particles are superpositions of two
distinctive bumps localized near external sources of both signs and
carrying fractional charge. In case of a half-integer charge of the
probe ($q=\frac{1}{2}$) the bumps carry charges $\frac{3}{4}$ and
$\frac{1}{4}$, so that the total probability is unity but the external
sources are completely screened, see fig.~\ref{wvfig}.

It is interesting that the screening effect would be probably not
easy to observe from Euclidean lattice simulations of the theory
(given by the partition function \rf{zed}). Indeed, the essence
of the mechanism is a formation of two bumps in the screening
wave functions, which is a kind of tunneling effect. The larger the
separation between sources $L$, the longer computer time one would need
for this effect to come into action, with the time growing
exponentially with $L$.

Nevertheless, it would be very useful to check the screening of
fractional charges by lattice simulations, in view of apparent
analogies with a more difficult case of non-Abelian gauge theories.

\subsection*{Acknowledgments}

D.D. would like to thank Victor Petrov for many discussions in the past
that stimulated this investigation.
K.Z. is grateful to NORDITA for hospitality while this work was
in progress. The work of K.Z. was supported in part by NATO Science
Fellowship, CRDF grant 96-RP1-253,
 INTAS grant 96-0524,
 RFFI grant 97-02-17927
 and grant 96-15-96455 for the promotion of scientific schools.
\vskip 2true cm

\setcounter{section}{0}
\appendix{Eigenfunction expansion of induced charge density}

Since the background field is time-independent, the eigenmodes
of the operator $-D^2-m^2$,
\be
(-D^2-m^2)\tilde{u}=\ka\tilde{u},
\ee
have the form
\be\label{ei}
\tilde{u}=\e^{i\es x^0}u,
\ee
where $u$ depends only on spatial coordinates and satisfies the
equation
\be\label{kg'}
\left[(\es+eA_0^{\rm cl})^2+\d_i\d_i-m^2\right]u_n=\ka_nu_n.
\ee
The eigenvalues $\ka_n$ are the functions of $\es$. The eigenfunctions
$u_n$ are supposed to be normalized to unity for any given $\es$.

The Green function \rf{green} can be expanded in the eigenfunctions as
\be\label{green1}
G(x,y)=\int_{-\infty}^{+\infty}\frac{d\es}{2\pi}\,\e^{i\es(x^0-y^0)}
\sum_n\frac{u_n(x)\bar{u}_n(y)}{\ka_n(\es)+i0}.
\ee
The integral over $\es$ can be calculated closing the contour of
integration in the upper or in the lower complex half-plane, depending
on the sign of $x^0-y^0$. The charge density is defined in the limit
$x^0\rightarrow y^0$ and, in principle, depends on how this limit is
approached. The correct prescription is to take the limit symmetrically
in $x^0-y^0$.

The integrand in \rf{green1} has a pole if $\ka_n(\es)=0$. The values of
$\es$ which satisfy this condition are determined exactly by equation
\rf{kg}. The only difference is in the
normalization of the functions $u_n$
and $\psi_n$\footnote{We use here slightly different
notations than
in the main text where we discriminated
between the negative- and the positive-frequency modes.}:
\be\label{psiu}
\psi_n=\frac{u_n}{\sqrt{|N_n|}},
\ee
\be
N_n=2\int d^3x\,\bar{u}_n(\es_n+eA_0^{\rm cl})u_n.
\ee

For sufficiently weak fields all poles lie on the real axis.
Near a given pole one has
\be\label{derpol}
\ka_n(\es)=N_{n}(\es-\es_n)+\ldots,
\ee
since, as it follows from eq.~\rf{kg'},
$$
\frac{d\ka_n}{d\es}=N_n.
$$
For weak fields $N_n$ is positive for positive $\es$ and negative for
negative ones. Therefore, the contour of integration in eq.~\rf{green1}
passes the poles on the positive semiaxis from above and the ones
\begin{figure}[t]
\hspace*{1cm}
\epsfysize=6cm
\epsfbox{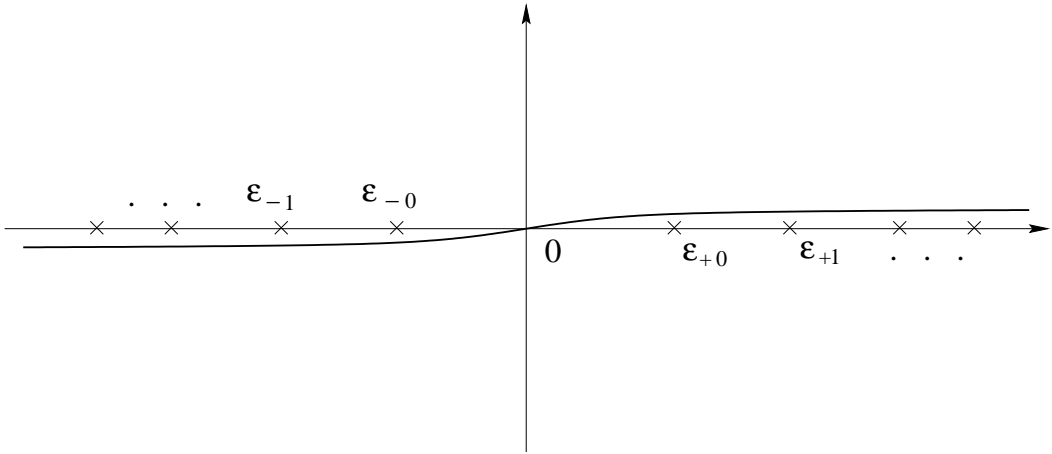}
\caption[x]{The contour of integration in the complex
$\protect\es$ plane.}
\label{cpfig}
\end{figure}
on the negative semiaxis from below (fig.~\ref{cpfig}). By continuity,
this rule remains valid for strong
fields until the positive and the negative
poles collide. It can be shown that
exactly when two poles collide
(fig.~\ref{specfig}) and go off to the complex plane
the integral $N_n$ turns to zero \cite{zp71}.

As a consequence of eqs.~\rf{derpol},~\rf{psiu}, the functions $u_n$
are replaced by $\psi_n$ in the residues of the integral \rf{green1}.
Taking into account that the covariant derivative $D_0$ acts on the
eigenfunctions \rf{ei} of the Klein-Gordon operator as
$i(\es+eA_0^{\rm cl})$,
we find for the induced charge density:
\be
\mn{J_0}=e\sum_n\bar{\psi}_n(\es_n+eA_0^{\rm cl})\psi_n.
\ee
It can be checked that this expression coincides with the vacuum
average of the charge density operator defined in sec.~\ref{hsec}.

The above consideration implies that
all eigenvalues of the Klein-Gordon equation lie on the
real axis, which is true for sufficiently weak fields, that is, for
sufficiently small separation between external charges $L$.
For very strong fields (at $L>L_0$)
some poles move to the complex plane, as discussed in
Sec.~\ref{kgsec}, which invalidates the analysis above. This signals
that for large $L$ the ground state is rearranged and the true vacuum
is different from an empty state with no dynamical charges.


\begin{thebibliography}{99}
\addtolength{\itemsep}{-6pt}

\bibitem{CJS71}
S. Coleman, R. Jackiw and L. Susskind, Ann. Phys. 93 (1971) 267.

\bibitem{GKMS95}
D.J. Gross, I.R. Klebanov, A.V. Matytsin and A.V. Smilga,
Nucl. Phys. B461 (1996) 109, hep-th/9511104.

\bibitem{ada96}
R. Rodriguez and Y. Hosotani, Phys. Lett. B375 (1996) 273,
hep-th/9602029;\\
C. Adam, Phys. Lett. B394 (1997) 161, hep-th/9609155.

\bibitem{YM2}
Y. Frishman and J. Sonnenschein, Nucl. Phys. B496 (1997) 285,
hep-th/9701140;\\
S. Dalley, Phys. Lett. B418 (1998) 160, hep-th/9708115;\\
A. Armoni, Y Frishman and J. Sonnenschein, Phys. Rev. Lett. 80 (1998) 430,
hep-th/9709097; hep-th/9807022.

\bibitem{KS95}
D. Kutasov and A. Schwimmer, Nucl. Phys. B442 (1995) 447,
hep-th/9501024. 

\bibitem{AB97}
E. Abdalla and R. Banerjee, Phys. Rev. Lett. 80 (1998) 238,
hep-th/9704176; \\
E. Abdalla, R. Banerjee and C. Molina, hep-th/9808003.

\bibitem{CS}
I. Affleck, J.A. Harvey and E. Witten, Nucl. Phys. B206 (1982) 413;\\
A.J. Niemi and G.W. Semenoff, Phys. Rev. lett. 51 (1983) 2077;\\
A.N. Redlich, Phys. Rev. D29 (1984) 2366.

\bibitem{mig72}
A.B. Migdal, Sov. Phys. JETP 34 (1972) 1184 [Zh. Eksp. Teor. Fiz.
61 (1972) 2209].

\bibitem{zp71}
Ya.B. Zel'dovich and V.S. Popov, Sov. Phys. Uspekhi 14 (1972)
673 [Usp. Fiz. Nauk 105 (1971) 403].

\bibitem{GMR85}
W. Greiner, B. M\"uller and J. Rafelski, {\it Quantum Electrodynamics
of Strong Fields} (Springer-Verlag, 1985).

\end{thebibliography}
\end{document}